# A spin-torque nano-oscillator based on interlayer-coupled meron-skyrmion pairs with a fixed orbit

Qiyun Yi,[1] Ting Han,[2] Jinyi Jiang,[3] and Xiangjun Xing[1,*]

[1]*School of Physics & Optoelectronic Engineering, Guangdong Provincial Key Laboratory of Information Photonics Technology, Guangdong University of Technology, Guangzhou 510006, China*

[2]*School of Materials Science & Engineering, Taiyuan University of Science and Technology, Taiyuan 030024, China*

[3]*School of Physics & Astronomy, Yunnan University, Kunming 650091, China*

**ABSTRACT:** In recent years, magnetic skyrmion-based spin-torque nano-oscillators (STNOs) attract considerable interest for their prospect in future-generation communication and spintronic technologies. However, some critical issues, which hamper their practical applications, e.g., the long start-up time and variable skyrmion gyration orbit, remain to be resolved. Here, we numerically demonstrate a realization of a fixed-orbit STNO, which is based on an interlayer-coupled meron-skyrmion (MS) pair other than a magnetic skyrmion. In this STNO, the MS pair possesses a structurally defined, fixed orbit within a broad range of driving current, even in the presence of random defects. The output frequency range of the STNO based on an MS pair far exceeds that of the STNO typically based on a single skyrmion. Moreover, the output frequency of this STNO can be further elevated if more MS pairs are incorporated. Our results reveal the nontrivial dynamics of the interlayer-coupled MS pair, opening perspectives for the design and optimization of fundamental spintronic devices.

**KEYWORDS:** spin-torque nano-oscillator, meron, skyrmion, fixed orbit, interlayer exchange coupling

---

[*]Email: xjxing@gdut.edu.cn

## I. INTRODUCTION

Magnetic skyrmions, as particle-like spin textures, hold enormous potential in developing information storage and diverse spintronic devices owing to their topologically protected stability, small size, and expected low driving current.[1–5] The topological charge $Q = \frac{1}{4\pi} \iint \mathbf{m} \cdot \left(\frac{\partial \mathbf{m}}{\partial x} \times \frac{\partial \mathbf{m}}{\partial y}\right) dx\, dy$ of a ferromagnetic (FM) skyrmion is equal to ±1, where $\mathbf{m} = \mathbf{m}(x, y)$ is the normalized magnetization vector.[5] Triggered by a sufficiently large driving force, a skyrmion—when touching the sample edge—loses the topological protection and converts into an edge-meron with a fractional topological charge,[6] which risks being expelled from the sample by an outward attractive force from the edge.[7,8] Conversely, the creation of magnetic skyrmions in a constricted geometry tends to begin from the nucleation of edge-merons.[9–11]

Spin-torque nano-oscillators (STNOs) function as microwave generators by converting an input direct current into a microwave voltage.[12] Benefiting from the nanoscale size and compatibility with existing CMOS technology, they are considered key ingredients of next-generation spintronic devices. As compared to other types of STNOs, such as the ones featuring a single domain[13–15] or magnetic vortex,[16–18] skyrmion-based STNOs offer several advantages, including the reduced driving current, smaller frequency linewidth, wider tunable frequency range, and synchronization of multiple skyrmions in a single STNO.[19,20] However, there exist a few innate shortcomings for skyrmion-based STNOs:[19,20] specifically, the frequency of skyrmion-based STNOs comprising a single FM nanodisk is typically restricted to a few gigahertz, the initiation time for the current-driven skyrmion to enter a steady orbit is undesirably long (typically, 20 ns), and the orbit of skyrmion gyration depends on the applied current complicating the design of detection units based on the magnetoresistance effect.

Replacing FM skyrmions with antiferromagnetic skyrmions or synthetic antiferromagnetic (SAF) skyrmion pairs can raise the frequency of STNOs naturally.[21–23] To reduce the start-up time, the skyrmion gyrational motion in an STNO can be harnessed by etching grooves and/or tailoring the magnetic properties in a ring area of the nanodisk.[24–28] Nevertheless, it remains challenging to accurately modify a unique ring area in a nanodisk using even state-of-the-art experimental techniques. Until now, a solution that can simultaneously address all the stated problems is still lacking.

By subtly combining distinct spin textures and exploiting their interactions, emerging functionalities and applications unattainable with individual magnetic entities might be opened. For example, a stripe domain wall can enhance the longitudinal motion of skyrmions in a nanotrack[29–31] and periodically spaced spin textures can act as reconfigurable magnonic crystals.[32–35] Recently, Deng et al.[36] conceptually proposed a skyrmion mirroring device with a hybrid tandem structure and numerically demonstrated the transition of a skyrmion into an interlayer-coupled skyrmion–skyrmion (SS) pair.

In this research, we devise an STNO based on an SAF nanodisk and exploit the gyrational dynamics of an interlayer-coupled meron–skyrmion (MS) pair stabilized in the SAF structure. We employ micromagnetic simulations to examine the spectral characteristics of the STNO dependent on the geometric and material properties. It is seen that the frequency range of the MS pair-based STNO is significantly extended relative to that of a typical skyrmion-based STNO. Increasing the number of MS pairs in the STNO can further elevate its output frequency. Moreover, weak disorder in the sample does not remarkably modify the main characteristics of the STNO.

## II. MODEL AND METHODS

As depicted in Fig. 1(a), the STNO consists of three main parts: the fixed layer, the spacer layer, and the free layer. Unlike in a conventional STNO, an SAF multilayer structure serves as the free layer in our STNO. The lower nanodisk with a bigger radius $r_l$ (layer-l) and the upper nanodisk with a smaller radius $r_u$ (layer-u), separated by a ruthenium (Ru) layer, form an asymmetric SAF structure, so that only the center of layer-l is subject to the Ruderman-Kittel-Kasuya-Yosida (RKKY) interlayer exchange coupling with layer-u. Both layer-l and layer-u assume an FM/HM (heavy-metal) bilayer structure to enable the presence of a Dzyaloshinskii-Moriya interaction (DMI) alongside the perpendicular magnetocrystalline anisotropy (PMA). The fixed layer harbors a radial vortex-like magnetization configuration.[20–24,26,27] An out-of-plane electric current is fed into the STNO through a top electrode with a radius $r_e$.

The public-domain micromagnetic solvers MuMax3[37] and OOMMF[38] (modified with the interfacial DMI extension module[39]) are employed to find the static magnetization configuration and subsequently to trace the current-induced dynamics of magnetization, which is governed by the Landau-Lifshitz-Gilbert-Slonczewski equation,[19,20]

$$\partial \mathbf{m}/\partial t = -\gamma(\mathbf{m} \times \mathbf{H}_{\text{eff}}) + \alpha(\mathbf{m} \times \partial \mathbf{m}/\partial t) + \mathbf{T}_s, \quad (1)$$

where $\gamma$ is the gyromagnetic ratio, $\alpha$ is the Gilbert damping constant, $t$ is time, and $\mathbf{H}_{\text{eff}}$ is the effective magnetic field, which includes the Heisenberg exchange field, magneto-dipolar field, perpendicular anisotropic field, DMI effective field, and interlayer exchange field. $\mathbf{T}_s = \gamma B_J(\mathbf{m} \times \mathbf{p} \times \mathbf{m})$ is the spin-transfer torque, where the spin-polarization vector has a radial vortex-like distribution, $\mathbf{p} = (\cos\varphi_p, \sin\varphi_p, 0)$, with $\varphi_p = \arctan\left(\frac{y-r_l}{x-r_l}\right) + \varphi$[20] and $B_J =$ $\frac{\hbar PJ}{2e\mu_0 M_s t_f}$ is the efficiency factor with $\hbar$ representing the reduced Planck constant, $P$ the spin polarization ratio, $J$ the current density, $\mu_0$ the vacuum permeability, $e$ the elementary charge, $M_s$ the saturation magnetization, and $t_f$ the thickness of each FM layer in the SAF nanodisk.

The material parameters for both free layers (layer-u and layer-l) feature the Pt/Co multilayer film:[40–42] the saturation magnetization $M_s$ = 580 kA/m, exchange stiffness $A$ = 15 pJ/m, PMA constant $K_u$ = 0.8 MJ/m$^3$, DMI strength $D$ = 3.5 mJ/m$^2$, and Gibert damping constant $\alpha$ = 0.02. The bilinear interfacial exchange coefficient quantifying the strength of interlayer antiferromagnetic exchange interaction is $\sigma$ = -2 mJ/m$^2$ unless explicitly stated otherwise.[43] It is assumed that $P$ = 0.4 and the spin-polarization attenuation between layer-u and layer-l is negligible, so that the spin-transfer torque manifests an identical efficiency in both free layers. This supposition is reasonable considering the small thickness of each layer in the SAF nanodisk.[44] The unit cell size 1×1×1 nm$^3$ is adopted regardless of the sample size. Open boundary conditions are assumed.

## III. RESULTS AND DISCUSSION

To create an MS pair, a spin-polarized current can be applied locally to a desired region at the boundary of layer-l, where the magnetization is switched under the action of spin-transfer torque. Subsequently, a mirroring nucleation site for the meron is formed in layer-u as a result of the strong interlayer exchange coupling.[36] After the nucleation of the spin textures, the applied current is turned off allowing the system to relax freely. During the relaxation process, the meron experiences an outward attractive force $\mathbf{F}_m^e$ from the edge of layer-u and an inward dragging force $\mathbf{F}_m^I$ from the skyrmion. Meanwhile, the skyrmion undergoes an inward repulsive force $\mathbf{F}_s^e$ from the edge of layer-l and an outward dragging force $\mathbf{F}_s^I$ from the meron.

In a certain range of $r_u/r_l$ [highlighted in Fig. 1(b) by the white box], the resultant force on the MS pair can become zero (specifically, $\mathbf{F}_m^e + \mathbf{F}_m^I = 0$ and $\mathbf{F}_s^e + \mathbf{F}_s^I = 0$), so that the interlayer-coupled MS pair is stabilized near the edge of layer-u. As revealed by the simulation results, the resulting equilibrium spin configuration carries a topological charge $Q$ ~ 0.5 [as illustrated in Fig. 1(c), middle panel], which is consistent with the expected value of the topological charge ($Q = Q_{em} + Q_{sk}$) of a pair of edge-meron ($Q_{em}$ = -0.5) and skyrmion ($Q_{sk}$ = +1). At a smaller $r_u/r_l$ than required [e.g., the region in Fig. 1(b) marked by the yellow box], the injected skyrmion is remote from the border of layer-l, whereby the inward force $\mathbf{F}_s^e$ on the skyrmion from the edge of layer-l cannot equilibrate the outward force $\mathbf{F}_m^e$ on the meron from the edge of layer-u. Consequently, the meron is expelled from layer-u and meanwhile the skyrmion is dragged toward the edge of layer-l. Ultimately, the skyrmion resides between the borders of layer-u and layer-l, because it would encounter a potential barrier when approaching either border.[36,45] For this configuration, the topological charge of the SAF structure is $Q \sim 1$ [Fig. 1(c), left panel]. When $r_u/r_l$ is much higher [e.g., the region in Fig. 1(b) indicated by the magenta box], both the meron and the skyrmion are expelled out, resulting in two antiferromagnetically coupled single domains with a topological charge $Q \sim 0$ [Fig. 1(c), right panel]. Note that no SS pair (also known as bilayer-skyrmion[40]) appears in the phase diagram for the considered values of $[r_u, r_l]$, since the formation of an SS pair involves conversion of the meron to a skyrmion—a process associated with a high potential barrier.

This study concentrates on the dynamics of an MS pair driven by a perpendicular current. Unless otherwise specified, all results and discussion are based on the STNO with $[r_u, r_l]$ = [40 nm, 60 nm] and $r_e$ = 40 nm. The snapshots of the MS pair under a given current density are depicted in Fig. 2(a), indicating that the meron and the skyrmion are tightly bound by the interlayer exchange interaction and move synchronously along the edge of layer-u. Fig. 2(b) shows oscillations of the position ($R_x$, $R_y$) and average magnetization component $m_z$ of the meron. It is evident that, immediately after the application of the current, the MS pair moves steadily in a fixed circular orbit, resulting in a very short start-up time. The Fourier spectra of $R_x$ and $m_z$, plotted in Fig 2(c), displays the frequency ($f$ = 8.53 GHz) of the STNO at $J$ = 1.5×10$^{11}$ A/m$^2$, and the corresponding mode pattern shown in the inset identifies the trajectory of the MS pair. Note that the multiple, equidistantly spaced high harmonics in the frequency spectra arise from the truncated waveform of $m_z(t)$.[46]

Our simulations of the spectral characteristics of the STNO suggest that the frequency of the STNO not only varies with the applied current density, but also depends on the geometric and material properties of the SAF nanodisk. As shown in Fig. 3(a), for various values of the antiferromagnetic coupling coefficient $\sigma$, each curve of $f(J)$ terminates at a unique critical current density, above which the meron and the skyrmion cannot be bound together, unless the interlayer exchange coupling is further enhanced. That is, the critical current density scales with the interlayer exchange coupling strength. For sufficiently strong interlayer exchange coupling, the output frequency at a given current density appears insensitive to the magnitude of $\sigma$. However, the polarizer angle $\varphi$ can dramatically modify the characteristics of the STNO, as shown in Fig. 3(b). Specifically, the threshold current density of the STNO changes non-monotonically with $\varphi$. Once $\varphi$ attains $\pi/3$, the threshold current density drops rapidly. Fig 3(c) shows the output frequency versus driving current density at $\varphi$ = 0 for different values of $r_e$. It is obvious that a smaller electrode contributes to a reduced output frequency and an elevated threshold current density.

For comparison, we conduct additional simulations on an STNO based on an SS pair, which is stabilized in a symmetric SAF nanodisk,[23,40] where the two free layers separated by the Ru layer are identical in size, namely, $r_u = r_l = 60$ nm. As shown in Fig. 3(d), the output frequency of the STNO based on an SS pair predominates at small current densities, but drops below the frequency of the STNO based on an MS pair when the current density becomes high enough. We also examined the dynamics of an STNO based on a single-layer skyrmion[20] (in this case, the free layer-u and the Ru layer are absent). Under a negative driving current, the single skyrmion in the STNO can move counterclockwise in a circular orbit, manifesting a frequency about hundreds of megahertz. Its spectral characteristics is plotted in the inset of Fig. 3(d). As expected, if a positive current is applied, the skyrmion will return to the disk center in a clockwise sense.[20,21]

The current-induced circular motion of an MS pair can be described by the Thiele equation.[1–5,7,19–29,31,42,45] As illustrated in Fig. 4(a), for the meron,

$$\mathbf{F}_m^M + \mathbf{F}_m^e + \mathbf{F}_m^I + \mathbf{F}_m^\alpha + \mathbf{F}_m^J = 0, \quad (2)$$

and for the skyrmion,

$$\mathbf{F}_s^M + \mathbf{F}_s^e + \mathbf{F}_s^I + \mathbf{F}_s^\alpha + \mathbf{F}_s^J = 0, \quad (3)$$

where the first to fifth terms represent the Magnus force, the edge-induced force, the internal force between the meron and the skyrmion, the dissipative force, and the driving force, respectively. The Magnus force $\mathbf{F}^M = C\mathbf{G} \times \mathbf{v}$, where $\mathbf{v}$ denotes the tangential velocity of the MS pair, $\mathbf{G} = 4\pi Q \hat{z}$ is the gyrovector with $\hat{z}$ denoting the unit vector along $+z$, and $C = \mu_0 M_s t_f / \gamma$. The edge-induced force $\mathbf{F}^e = -\nabla U$, where U is the potential energy. The dissipative force $\mathbf{F}^\alpha = \alpha C \overleftrightarrow{\mathcal{D}} \cdot \mathbf{v}$, where $\overleftrightarrow{\mathcal{D}} = \begin{pmatrix} \mathcal{D}_{xx} & 0 \\ 0 & \mathcal{D}_{yy} \end{pmatrix}$ in which $\mathcal{D}_{xx} = \mathcal{D}_{yy} \equiv \mathcal{D} = \int dxdy \left( \frac{\partial \mathbf{m}}{\partial i} \cdot \frac{\partial \mathbf{m}}{\partial i} \right)$ with $i = (x, y)$. The current-induced driving force $\mathbf{F}^J = F^{Jr}\hat{\mathbf{e}}_r + F^{J\tau}\hat{\mathbf{e}}_\tau$ can be decomposed into a radial force and a tangential force, where $F^{Jj} = \gamma C B_J S$ in which $S = \int dxdy \left( (\mathbf{m} \times \mathbf{p}) \cdot \frac{\partial \mathbf{m}}{\partial j} \right)$ with $j = (r, \tau)$ denoting the radial or tangential direction. The internal force $\mathbf{F}^I = F^{Ir}\hat{\mathbf{e}}_r + F^{I\tau}\hat{\mathbf{e}}_\tau$ between the paired meron and skyrmion originates from the interlayer exchange coupling.

By adding Eq. (2) to Eq. (3), one obtains the Thiele equation for an MS pair in steady motion, $\mathbf{F}^M + \mathbf{F}^e + \mathbf{F}^\alpha + \mathbf{F}^J = 0$, where $\mathbf{F}^I$ as an internal force is absent. As illustrated in Fig.

4(a), these forces can be projected onto the radial and tangential directions. In the circular motion, the radial velocity of the MS pair is zero, and the net velocity of the MS pair can be derived from the Thiele equation in the tangential direction, $\mathbf{F}^\alpha + \mathbf{F}^J = 0$, i.e., $\alpha \mathcal{D} v \hat{\mathbf{e}}_\tau - \gamma S B_J \hat{\mathbf{e}}_\tau = 0$. To simply the calculation, we consider an instant when the MS pair is passing through the rightmost end of the circular orbit, where the velocity of the MS pair is oriented along -$y$. In this scenario, $\mathcal{D} = \int \mathrm{d}x \mathrm{d}y \left( \frac{\partial \mathbf{m}}{\partial y} \cdot \frac{\partial \mathbf{m}}{\partial y} \right)$ and $S = \int \mathrm{d}x \mathrm{d}y \left( (\mathbf{m} \times \mathbf{p}) \cdot \frac{\partial \mathbf{m}}{\partial y} \right)$.

One will be able to calculate $\mathcal{D}$ and $S$ if an explicit form of $\mathbf{m}(x, y)$ is known. In fact, more than one ansatz of $\mathbf{m}(x, y)$ has been proposed for a skyrmion,[42,47,48] yet no proper ansatz exists for an edge-meron. Despite the lack of an available ansatz for the meron, numerical values of $\mathcal{D}$ and $S$ can still be found from the simulation results by rewriting $\frac{\partial \mathbf{m}}{\partial y}$ in the discrete form, $\frac{\partial \mathbf{m}}{\partial y} \rightarrow \frac{\Delta \mathbf{m}}{\Delta y} = \frac{\mathbf{m}(y+\Delta y) - \mathbf{m}(y-\Delta y)}{2\Delta y}$. Then, after some procedures, we obtain the numerical values of $\mathcal{D}$ and $S$ for the meron and the skyrmion. Considering specifically an MS pair at $J = 5 \times 10^{10}$ A/m$^2$, for the meron, $\mathcal{D}_m = 95.0$ and $S_m = 93.9$ nm, while for the skyrmion, $\mathcal{D}_s = 172.3$ and $S_s = 96.7$ nm. Having these numerical values, one acquires the velocity of the MS pair, $\mathbf{v} = \frac{\gamma S_p}{\alpha \mathcal{D}_p} B_J \hat{\mathbf{e}}_\tau$, where $\mathcal{D}_p = \mathcal{D}_m + \mathcal{D}_s$ and $S_p = S_m + S_s$. The dependencies of $\mathcal{D}$ and $S$ upon the current density can be found in Fig. 5.

As revealed by the data for the MS pair, the value of $\mathcal{D}$ for the meron is far smaller than that of the skyrmion, whereas the values of $S$ for both the meron and the skyrmion are almost equal. In all cases, $S$ is more sensitive to the current density than $\mathcal{D}$. In particular, for the MS pair, $S$ rises with the increased current density, whereas for the SS pair and the single-layer skyrmion, $S$ diminishes with the increased current density. These features result in the consequence that, at a sufficiently high current density, the MS pair moves faster and possesses a higher frequency than the SS pair. Another important feature is that, in the relevant range of current density, the value of $S$ for the single-layer skyrmion is several times smaller than those of the MS and SS pairs, which explains why an STNO based on a single skyrmion has very small frequencies.[19,20]

To identify the role of interlayer exchange interaction in the synchronized motion of the MS pair, we carry out a set of control simulations, where the spin-polarized current only flows into a single free layer, either layer-u or layer-l, at a time.[22] In this scenario, both frequencies of the MS pair and the SS pair are seen to drop [Figs. 3(d) and 4(b)], implying that a double-layer current is preferable in order to synchronize the meron and the skyrmion.[22,49]

One distinctive feature of the skyrmion-based STNO lies in the synchronized motion of multiple skyrmions in a single nanodisk, which can elevate the output frequency remarkably.[19,23] Following this strategy, we arrange multiple MS pairs ($N$ = 2, 3, 4, 5) in the SAF nanodisk, as illustrated in Fig. 6(a), and investigate their dynamics under a spin-polarized current. As anticipated, the frequency of the STNO with multiple MS pairs increases with the increasing number $N$, as shown in Fig. 6(b). It is worthy to note that, the symmetric distribution of the MS pairs would be destroyed if an exceedingly large current is utilized, leading to the emergence of additional frequencies.

Finally, we examine the influence of disorder on the dynamic characteristics of an MS pair. Typically, we concentrate on a kind of disorder closely related to the polycrystalline ultrathin films fabricated by magnetron sputtering.[31,50] To capture the key property of a granular film, we introduce the random variation $\Delta K_u$ to the PMA constant $K_u$ and employ Voronoi tessellation to generate two grain patterns: i) the mean grain size is 10 nm and $\Delta K_u$ from grain to grain is within 5%, as depicted in Fig. 7(a), and ii) the average grain size is 25 nm and $\Delta K_u$ is within 2%, as indicated in Fig. 7(b). The frequency spectra and mode patterns of an MS pair in the granular and clean samples, plotted in Fig. 7(c), suggest that the weak disorder only leads to a fluctuating velocity of the MS pair but has no appreciable effect on its gyro-orbit.

**IV. CONCLUSION**

In summary, we propose a spin-torque nano-oscillator (STNO) that is based on an interlayer-coupled meron–skyrmion (MS) pair. The output characteristics of the STNO has been examined via micromagnetic simulations and the Thiele model. Comparative simulations reveal that this STNO surpasses, in several aspects, the STNO based on either interlayer-coupled skyrmion–skyrmion pair or conventional single-layer skyrmion. The synchronized motion of multiple MS pairs in a single STNO can raise the output frequency of the STNO further by several times. Furthermore, the presence of weak disorder appears not to markedly modify the key characteristics of the STNO. This type of STNO holds great potential to overcome some critical bottlenecks, such as the ultralong initiation time, experienced by skyrmion and vortex-based STNOs in practical applications. Our research suggests that the interplay of distinct features[29,31,51] provides a viable route toward settling some critical technical issues.

**ACKNOWLEDGEMENTS:** X.X. acknowledges support by the Guangdong Basic and Applied Basic Research Foundation under Grant No. 2022A1515010605 and the National Natural Science Foundation of China under Grant No. 11774069. Author Contributions: X.X. initiated the study. Q.Y. and J.J. designed the meron-skyrmion pair. Q.Y. performed numerical simulations. X.X. developed the Thiele model. All authors discussed the results. Q.Y. and X.X. wrote the manuscript with input from T.H. and J.J.

**Competing Interests:** The authors declare that they have no competing interests.

**Data Availability:** The data that support the findings of this study are available from the corresponding author upon reasonable request.

**REFERENCES**

[1]A. Fert, N. Reyren, and V. Cros, Magnetic Skyrmions: Advances in Physics and Potential Applications, Nat. Rev. Mater. **2**, 17031 (2017).

[2]W. Jiang, G. Chen, K. Liu, J. Zang, S. G. E. te Velthuis, and A. Hoffmann, Skyrmions in Magnetic Multilayers, Phys. Rep. **704**, 1 (2017).

[3]B. Göbel, I. Mertig, and O. A. Tretiakov, Beyond Skyrmions: Review and Perspectives of Alternative Magnetic Quasiparticles, Phys. Rep. **895**, 1 (2021).

[4]H. Yu, J. Xiao, and H. Schultheiss, Magnetic Texture Based Magnonics, Phys. Rep. **905**, 1 (2021).

[5]N. Nagaosa and Y. Tokura, Topological Properties and Dynamics of Magnetic Skyrmions, Nat. Nanotechnol. **8**, 899 (2013).

[6]M. Pereiro, D. Yudin, J. Chico, C. Etz, O. Eriksson, and A. Bergman, Topological Excitations in a Kagome Magnet, Nat. Commun. **5**, 4815 (2014).

[7]X. Xing, P. W. T. Pong, and Y. Zhou, Current-Controlled Unidirectional Edge-Meron Motion, J. Appl. Phys. **120**, 203903 (2016).

[8]J.-P. Chen, J.-Q. Lin, X. Song, Y. Chen, Z.-F. Chen, W.-A. Li, M.-H. Qin, Z.-P. Hou, X.-S. Gao, and J.-M. Liu, Control of Néel-Type Magnetic Kinks Confined in a Square Nanostructure by Spin-Polarized Currents, Front. Phys. **9**, 680698 (2021).

[9]J. Iwasaki, M. Mochizuki, and N. Nagaosa, Current-Induced Skyrmion Dynamics in Constricted Geometries, Nat. Nanotechnol. **8**, 742 (2013).

[10]Y. Zhou and M. Ezawa, A Reversible Conversion between a Skyrmion and a Domain-Wall Pair in a Junction Geometry, Nat. Commun. **5**, 4652 (2014).

[11]W. Jiang, P. Upadhyaya, W. Zhang, G. Yu, M. B. Jungfleisch, F. Y. Fradin, J. E. Pearson, Y. Tserkovnyak, K. L. Wang, O. Heinonen, S. G. E. te Velthuis, and A. Hoffmann, Blowing Magnetic Skyrmion Bubbles, Science **349**, 283 (2015).

[12]T. Chen, R. K. Dumas, A. Eklund, P. K. Muduli, A. Houshang, A. A. Awad, P. Dürrenfeld, B. G. Malm, A. Rusu, and J. Åkerman, Spin-Torque and Spin-Hall Nano-Oscillators, Proc. IEEE **104**, 1919 (2016).

[13]S. I. Kiselev, J. C. Sankey, I. N. Krivorotov, N. C. Emley, R. J. Schoelkopf, R. A. Buhrman, and D. C. Ralph, Microwave Oscillations of a Nanomagnet Driven by a Spin-Polarized Current,
10

**Figures captions**

**Figure 1.** Device structure and possible spin configurations. (a) Schematic diagram of a spin-torque nano-oscillator (STNO). The STNO includes three functional layers: the fixed layer, the spacer layer, and the free layer. The free layer comprises a synthetic antiferromagnetic (SAF) nanodisk, in which the upper free layer (layer-u) and the lower free layer (layer-l) are separated by an ultrathin ruthenium (Ru) layer to ensure the occurrence of interlayer exchange coupling. Both layer-u and layer-l are supposed to adopt a ferromagnet/heave-metal (FM/HM) structure to stabilize chiral spin textures by introducing a Dyzaloshinskii-Moriya interaction. The radii of layer-u and layer-l are denoted as $r_u$ and $r_l$, respectively. Layer-l is set to be wider than layer-u, i.e., $r_l > r_u$, to facilitate the nucleation of a meron-skyrmion (MS) pair. The fixed layer assumes a radial vortex-like magnetization distribution. The current flows perpendicular to the fixed layer via a top electrode with a radius $r_e$ (not shown for brevity). Without external stimuli, the outward attractive force $\mathbf{F}_m^e$ on the meron from the edge of layer-u counteracts the inward repulsive force $\mathbf{F}_s^e$ on the skyrmion from the edge of layer-l, via the mediation of the internal forces $\mathbf{F}_m^I$ and $\mathbf{F}_s^I$ between the meron and the skyrmion, resulting in stabilization of the MS pair at the border of layer-u. The origin of coordinate is at the lower-left rim of layer-l. (b) Phase diagram for the spin configurations in the SAF nanodisk. This diagram, rendered as dependency of the total topological charge in the SAF nanodisk on the radii $r_u$ and $r_l$, contains three regions, typically, the yellow, white, and magenta dashed boxes highlighting $Q = 0.93$ for $[r_u, r_l] = [24$ nm, $60$ nm$]$, $Q = 0.52$ for $[r_u, r_l] = [40$ nm, $60$ nm$]$, and $Q = -0.04$ for $[r_u, r_l] = [56$ nm, $60$ nm$]$, respectively. (c) The spin configurations in the SAF nanodisk corresponding to $Q = 0.93$, $Q = 0.52$, and $Q = -0.04$. The spin configuration with $Q = 0.52$ represents an MS pair.

**Figure 2.** Frequency characteristics of an MS pair. (a) Temporal sequences of the MS pair under a current, illustrating clockwise gyration. The current density $J = 1.5 \times 10^{11}$ A/m$^2$. The orange grid box in layer-u highlights the detection region of the output signal. (b) Oscillations of the positions ($R_x$ and $R_y$) and average magnetization component $m_z$ of the meron with time. (c) Fourier-transform spectra of $R_x$ and $m_z$. The inset shows the mode pattern of the gyrating MS pair with a frequency of $f = 8.53$ GHz. In these illustrations, $[r_u, r_l] = [40$ nm, $60$ nm$]$. The output signal $m_z$ is detected, at the region marked by the orange grid box in each panel, via the magnetoresistance effect.

**Figure 3.** Spectral characteristics of an MS pair. Panels (a)-(d) plot the frequency versus current density for various parameters: (a) strengths $\sigma$ of the interlayer exchange coupling, (b) polarizer angle $\varphi$, (c) radius $r_e$ of the electrode, and (d) radius $r_u$ of layer-u.

**Figure 4.** Force diagram and coupling-assisted motion of an MS pair. (a) Thiele forces on the meron and the skyrmion in steady motion. The white arrowed circle indicates the direction of motion of the MS pair. In the radial direction, for both the meron and the skyrmion, the radial component of the internal force ($\mathbf{F}_m^{Ir}$, $\mathbf{F}_s^{Ir}$) between the meron and the skyrmion cancels off the Magnus force ($\mathbf{F}_m^M$, $\mathbf{F}_s^M$) and the edge-induced force ($\mathbf{F}_m^e$, $\mathbf{F}_s^e$). In the tangential direction, for the meron, the current-induced force $\mathbf{F}_m^J$ is balanced by the dissipative force $\mathbf{F}_m^\alpha$ and the tangential component of the internal force $\mathbf{F}_m^{I\tau}$, whereas for the skyrmion, the dissipative force $\mathbf{F}_s^\alpha$ counteracts the current-induced force $\mathbf{F}_s^J$ and the tangential component of the internal force $\mathbf{F}_s^{I\tau}$. (b) Frequency versus current density for the MS pair (red triangles) and a skyrmion–skyrmion pair (blue triangles), with the current being selectively applied to layer-u (upright triangles) or layer-l (inverted triangles) at a time. In this scenario, the magnetic entity—in the layer without undergoing a current—moves alongside the other entity directly driven by a current, as a consequence of the internal forces $\mathbf{F}_m^{I\tau}$ and $\mathbf{F}_s^{I\tau}$ originating from the interlayer exchange coupling.

**Figure 5.** $\mathcal{D}$ and $S$ versus the current density for different magnetic entities: (a)(b) An MS pair in an SAF nanodisk with [$r_u$, $r_l$] = [40 nm, 60 nm], (c)(d) an SS pair in an SAF nanodisk with [$r_u$, $r_l$] = [60 nm, 60 nm], and (e)(f) a skyrmion in an FM nanodisk with [$r_u$, $r_l$] = [0 nm, 60 nm]. For all magnetic entities, $\mathcal{D}$ is insensitive to the current density, but the meron has a much smaller $\mathcal{D}$ than the skyrmion. While the values of $S$ for the MS pair rise with the current density, the values of $S$ for the SS pair and the single skyrmion decrease.

**Figure 6.** Spectral characteristics of an STNO carrying multiple MS pairs. (a) Centrosymmetrically distributed MS pairs in a single STNO. The number of the MS pairs, denoted as $N$, is indicated in each panel. In these illustrations, [$r_u$, $r_l$] = [56 nm, 76 nm]. (b) Comparison of the frequency versus current density for different numbers of MS pairs. Inset: zoom-in view of the curves at small current densities.

**Figure 7.** Effect of disorder on the characteristics of an MS pair. Panels (a) and (b) depict two realizations of the disorder in the SAF nanodisk modeled as a granular sample, in which both the grain size and the anisotropy constant $K_u$ are randomly distributed around their respective mean values. In these illustrations, [$r_u$, $r_l$] = [40 nm, 60 nm]. (a) The average grain size is 10 nm and $K_u$ in each grain varies randomly within 5% of the mean value $K_u^0$ = 0.8 MJ/m$^3$, namely, $\Delta K_u/K_u^0 = (K_u - K_u^0)/K_u^0 \leq 5\%$. (b) The average grain size is 25 nm and $K_u$ in each grain varies randomly within 2%, i.e., $\Delta K_u/K_u^0 \leq 2\%$. (c) Comparison of the frequency characteristics of an MS pair in clean and granular nanodisks under $J$ = 1.5×10$^{11}$ A/m$^2$. Insets: mode patterns of the STNO for the two granular samples. Apparently, the weak disorder in a granular nanodisk only causes a slight shift in the gyro-frequency of the MS pair and is not observed to significantly modify its orbit.

**Fig. 1**

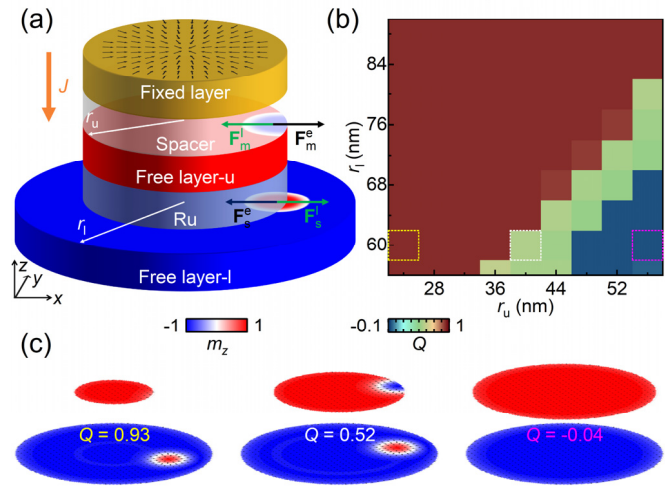

**Fig. 2**

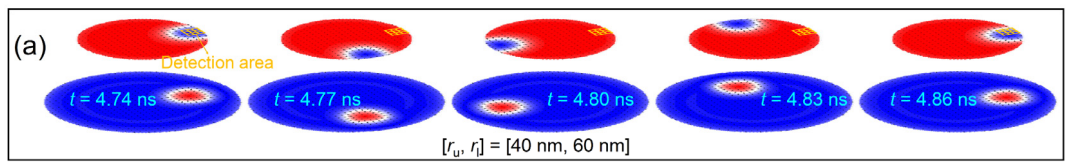

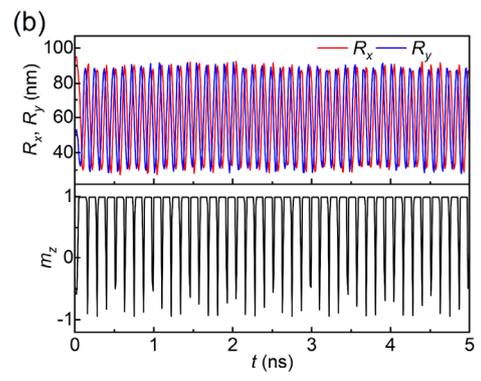
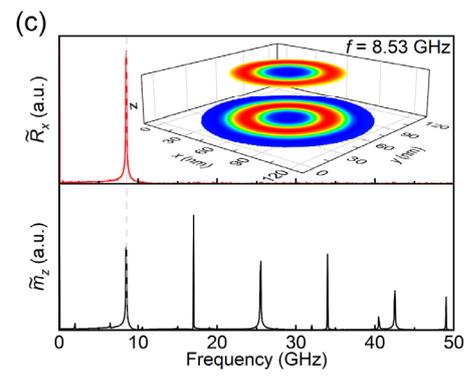

**Fig. 3**

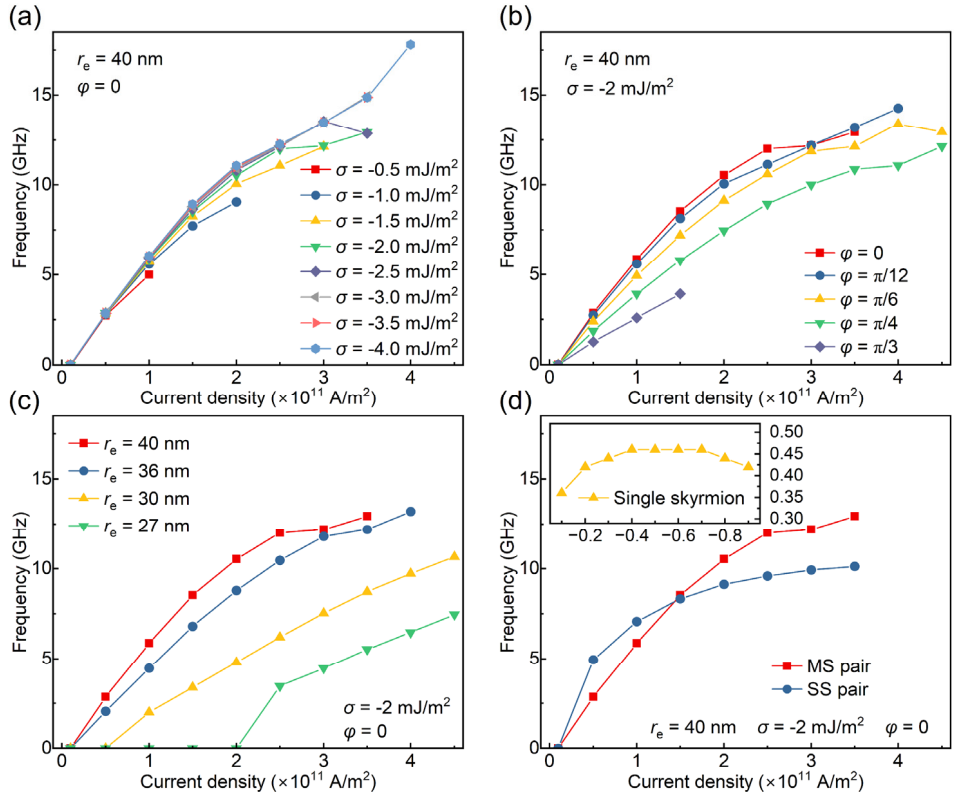
20

**Fig. 4**

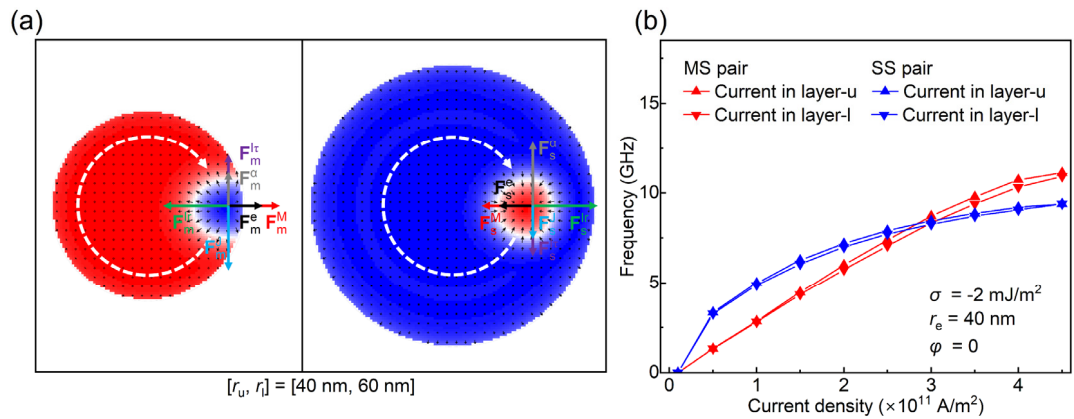

**Fig. 5**

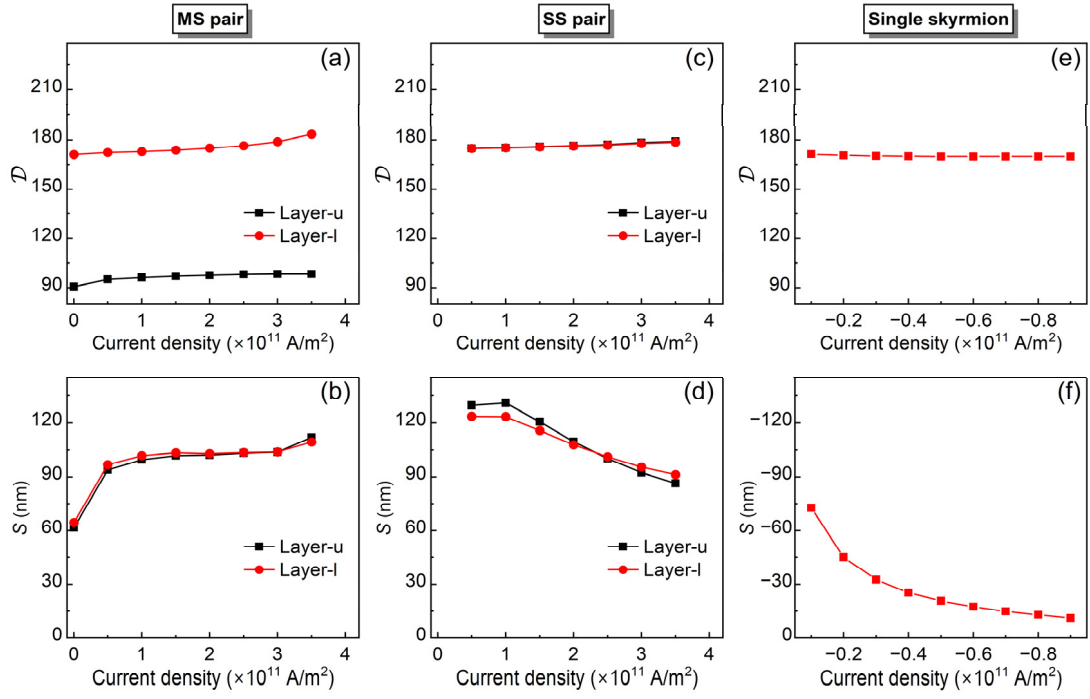

**Fig. 6**

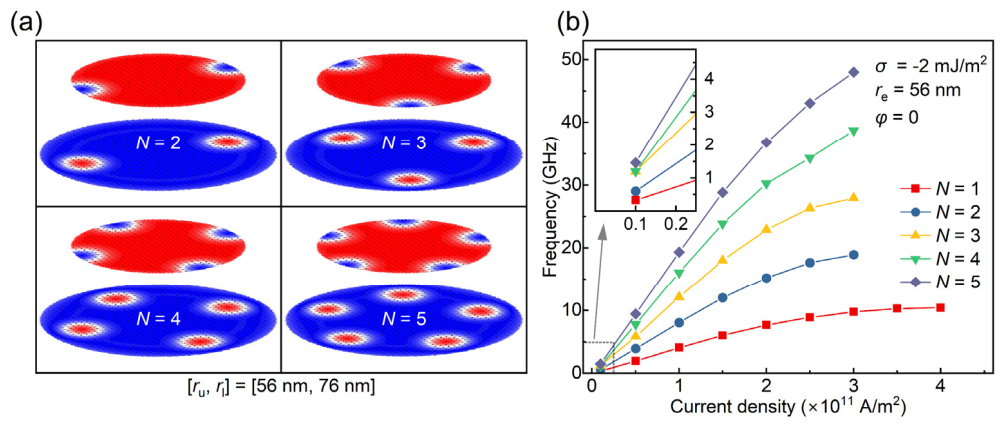

**Fig. 7**

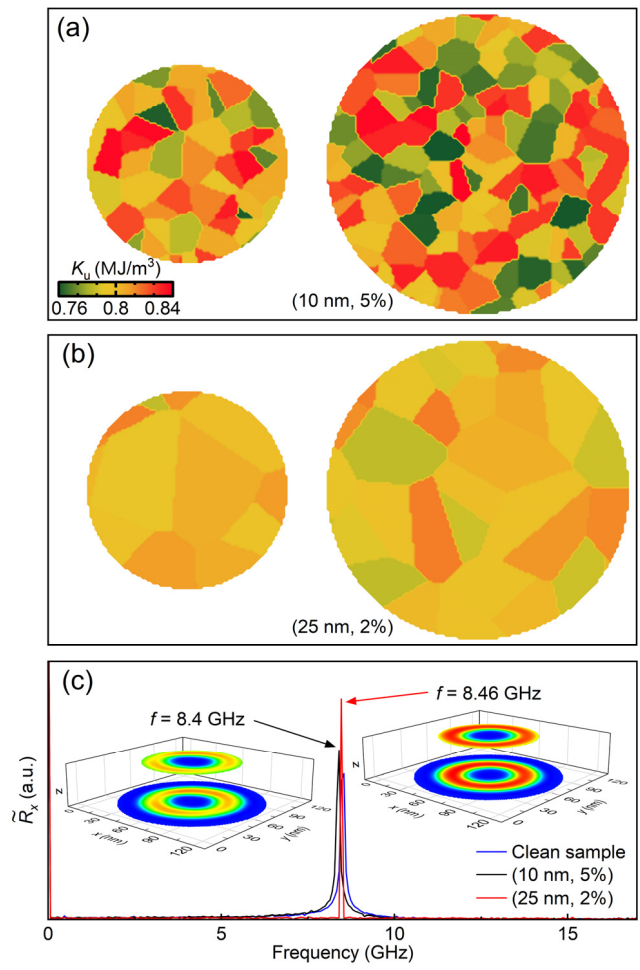